\RequirePackage{fix-cm}
\documentclass[11pt]{article}

\usepackage{graphicx}

\usepackage{listings}
\usepackage{color}

\definecolor{dkgreen}{rgb}{0,0.6,0}
\definecolor{gray}{rgb}{0.5,0.5,0.5}
\definecolor{mauve}{rgb}{0.58,0,0.82}

\usepackage{url}

\usepackage{breakurl}

\usepackage{bm}

\RequirePackage[OT1]{fontenc}
\RequirePackage{amsmath}
\RequirePackage[numbers]{natbib}

\RequirePackage[colorlinks,citecolor=blue,urlcolor=blue]{hyperref}

\usepackage{graphicx}
\usepackage{amsmath}
\usepackage{graphics}
\usepackage{epsfig}
\usepackage{amssymb}
\usepackage{color}
\usepackage{float}
\usepackage[section]{placeins}
\usepackage{subcaption,graphicx}

\usepackage[a4paper,bindingoffset=0.2in,%
            left=0.5in,right=0.5in,top=0.5in,bottom=0.8in,%
            footskip=.25in]{geometry}

\newcommand{\be}{\begin{eqnarray}}
\newcommand{\ee}{\end{eqnarray}}
\newcommand{\bea}{\begin{eqnarray*}}
\newcommand{\eea}{\end{eqnarray*}}


\begin{document}

\title{Generic probabilistic modelling and non-homogeneity issues \\for the UK epidemic of COVID-19
}

\author{ Anatoly Zhigljavsky\footnote{Cardiff University} , Roger Whitaker$^*$, Ivan Fesenko\footnote{University of Nottingham}, Kobi Kremnizer\footnote{University of Oxford},  Jack Noonan$^*$, \\
Paul Harper$^*$, Jonathan Gillard$^*$, Thomas Woolley$^*$,  Daniel Gartner$^*$,
 Jasmine Grimsley\footnote{ONS Data Science, Newport}\,, \\ Edilson de Arruda\footnote{
Federal University of Rio de Janeiro, Brazil}\,,
 Val Fedorov\footnote{Icon Clinical Research, Philadelphia, USA }\,, Tom Crick MBE\footnote{Swansea University}
}

\maketitle

\abstract{

Coronavirus COVID-19 spreads through the population mostly based on social contact. To gauge the potential for widespread contagion, to cope with associated uncertainty and to inform its mitigation,  more accurate and robust modelling is centrally important for policy making.

We provide a flexible modelling approach that increases the accuracy with which insights can be made. We use this to analyse different scenarios relevant to the COVID-19 situation in the UK. We present a stochastic model that captures the inherently probabilistic nature of contagion between population members. The computational nature of our model means that spatial constraints (e.g., communities and regions), the susceptibility of different age groups and other factors such as medical pre-histories can be incorporated with ease. We analyse different possible scenarios  of the COVID-19 situation in the UK. Our model is robust to small changes in the parameters and is flexible in being able to deal with different scenarios.

This approach goes beyond the convention of representing the spread of an epidemic through a fixed cycle of susceptibility, infection and recovery (SIR). It is important to emphasise that standard SIR-type models, unlike our model, are not flexible enough and are also not stochastic and hence should be used with extreme caution. Our model  allows both heterogeneity and inherent uncertainty to be incorporated. Due to the scarcity of verified data, we draw insights by calibrating our model using parameters from other relevant sources, including agreement on average (mean field) with parameters in SIR-based models.

 We use the model to assess parameter sensitivity for a number of key variables that characterise the COVID-19 epidemic. We also test several control parameters with respect to their influence on the severity of the outbreak.
{Our analysis shows that due to inclusion of spatial heterogeneity in the population and the asynchronous timing of the epidemic across different areas, the severity of the epidemic might be lower than expected from other models.}

{{We find that one of the most crucial control parameters that may significantly reduce the severity of the epidemic is the degree of separation of vulnerable people and people aged 70 years and over,
but note also that isolation of other groups has an effect on the severity of the epidemic. It is important to remember that models are there to advise and not to replace reality, and that any action
should be coordinated and approved by public health experts with experience in dealing with epidemics.}}

The computational approach makes it possible for further extensive scenario-based analysis to be undertaken. This and a comprehensive study of sensitivity of the model to different parameters defining COVID-19 and its development will be the subject of our forthcoming  paper. In that paper, we shall also extend the model where we will consider different probabilistic scenarios for infected people with mild and severe cases.

}

\newpage

\section{Introduction}

We model the development of the COVID-19  epidemic in the UK under different  scenarios of handling the epidemic.
There are many standard epidemiological models for modelling epidemics, see e.g. \cite{Daley,Hethcote}. In this research, we use a more generic simulation model which has the following features:

\begin{itemize}
  \item it can take into account spatial heterogeneity of the population and heterogeneity of development of  epidemic in different areas;
  \item it allows  the use of time-dependent strategies for { analysing} the epidemics;
   \item it allows taking into account special characteristics of particular groups of people, especially people with specific medical pre-histories and elderly.
\end{itemize}

Standard epidemiological models such as SIR and many of its modifications do not possess these properties and hence are not applicable for any study that requires the use of the features above.
In particular,
In particular, for influenza the mortality changes less significantly with age in comparison to coronavirus; hence common influenza models do not give much insight in modelling the COVID-19 epidemic. The report \cite{Ferguson}, which is widely considered as the main document specifying the current epidemic in the UK and in the world, is  almost entirely based on the use of standard epidemiological models and hence the conclusions of \cite{Ferguson} seem to be lacking specifics related to important issues of the study such as heterogeneity of development of  epidemic at different locations and even flexible use of different death rates across different ages. 
Unreliability of COVID-19 data, including the numbers of COVID cases and COVID deaths, is a serious problem. It is  discussed, in particular,
by G. Antes,  in \cite{Germ2}.

Despite the SIR models including the model of \cite{Ferguson} have been heavily criticised \cite{Shen},
  when choosing the main parameters of our model we calibrate it so that its mean field version approximately  reproduces the same output
as SIR-based models of \cite{Ferguson,Lourenco} and we use the parameter values consistent with~ \cite{Ferguson,Lourenco}. Also, the main notation (which sometimes does not look natural from a statistician's point of view) is taken from~\cite{Lourenco}.

The primary objective of our work  is  construction of a reliable, robust and interpretable  model describing  the epidemic  under different control regimes.  In this paper, we make the first step towards this objective. Due to lack of time and unreliability of available data on COVID-19, most of our results serve as   an illustration of the role of different control options and we hope that even as illustration they are  useful. However, we try to stay close to the COVID-19 epidemic scenario and hence  we use appropriate recommendations about the choice of parameters and models of the virus behaviour  we found from the studies based on the use of standard models.

See conclusions in Sections 3,4,5 for the main conclusions of this work.

\medskip

{{{\em Acknowledgement}.  We are grateful to several colleagues in Univ. Cambridge and Univ. Warwick for useful comments on drafts of this paper and to M. Hairer (ICL) and J. Ball (Univ. Nottingham) for  valuable comments and discussions. }}

\section{The  model}

\label{sec2}

\subsection*{ Variables and parameters}

\begin{itemize}
\item $t$ - time (in days)
\item $t_0$ - intervention time (e.g., the time when self-isolation starts)
  \item $N$ - population size
  \item $I(t)$ - number of infected at time $t$
  \item $R(t)$ - number of immune (recovered or dead) at time $t$ (this makes full sense from a modelling perspective, but we appreciate that this may look kind of strange when read by non-scientists)
  \item $S(t)=N-I(t)-R(t)$ - number of susceptible at time $t$
  \item $M$ - number of groups selected for the study
   \item $N_m$ -  size of $m$-th  group ($N=N_1+N_2+ \ldots + N_M$)
  \item $I_m(t)$ - number of infected at time $t$ in group $m$
  \item $R_m(t)$ - number of immune at time $t$ in group $m$
  \item $S_m(t)$ - number of susceptible at time $t$ in group $m$
   \item $\beta$ - average number of transmissions of the virus per unit of time with no intervention

  \item $1/\sigma$ - average infectious period
  \item $k$ - shape parameter of the Erlang distribution
   \item $R_0=\beta/\sigma$  - reproductive number (average number of people who will capture the disease from one contagious  person)
         \item  $R_{0}^\prime$  - reproductive number  after intervention
           \item $p_m$ - probability to recover for an infected person from the $m$-th group
\end{itemize}

\subsection*{ Values of parameters and generic model}

The  reproductive number $R_0$  is the main parameter defining the speed of development of an epidemic.
There is no true value for  $R_0$ as it varies  in different parts of the UK (and the world).
In particular, in rural areas one would expect a considerably lower value of $R_0$ than in London.
 Authors of \cite{Ferguson} suggest $R_0=2.2$ and $R_0=2.4 $ as typical; the authors of \cite{Lourenco} use values for $R_0$ in the range [2.25, 2.75].
We shall use the value $R_0=2.5$ as typical which may be a slightly pessimistic  choice  overall but  could be an adequate choice for the mega-cities where the epidemics develop faster and may lead to more causalities. In rural areas, in small towns, and everywhere else where social contacts are less intense,  the epidemic is milder.

We assume that the person becomes infected $\tau$ days after catching the virus, where $\tau$ has Poisson distribution with mean of 1 week. To model the time to recover (or die) we use  {\em Erlang distribution} with shape parameter $k=3$ and rate parameter $\lambda=1/7\,$ so that the mean of the distribution is $k/\lambda= 1/\sigma=21$ (in simulations, we discretise the numbers to their nearest integers). This implies that we assume that the average longevity of the period of time while the infected person is contagious is 21 days, in line  with the current knowledge, see e.g. \cite{Tang,Kucharski,womenshealth}. Standard deviation of the chosen Erlang distribution is approximately 12, which is rather large and reflects the uncertainty we currently have about the period of time a person needs to recover (or die) from COVID-19. An increase in $1/\sigma$ would prolong the epidemic and smaller values of  $1/\sigma$ would make it to cause people to be contagious for less time.  The use of Erlang distribution is standard for modelling similar events in reliability and queuing theories, which have much in common with epidemiology. We
have considered the sensitivity of the model in this
 study with respect to the choice
of parameters $\lambda$ and $\sigma$  but more has to be done in cooperation with epidemiologists. As there are currently many outbreaks epidemics, new knowledge about the distribution of the period of infection by COVID-19 can emerge soon.


The model varies depending on purposes of the study. The main ingredients are:  $I_m(t)$
is the birth-and-death processes and $R_m(t)$ is the associated pure birth processes.
The process of transmission is the Poisson process with intensity $\beta$ (time to next transmission has the exponential density $\beta e^{-\beta t}, \;t>0$). After the intervention (for $t \geq t_0$), the Poisson process of transmission for $m$-the group has  intensity $\beta_m$.

We treat this model as purely stochastic despite  parts of it can be written it terms of systems of stochastic differential equations. Despite running pure simulation models taking longer than running combined models, they are simpler and less prone to certain errors.

\subsection*{The number of infected at time $t$ as the main quantity of interest}

\begin{figure}[h]
\centering
  \includegraphics[width=0.85\linewidth]{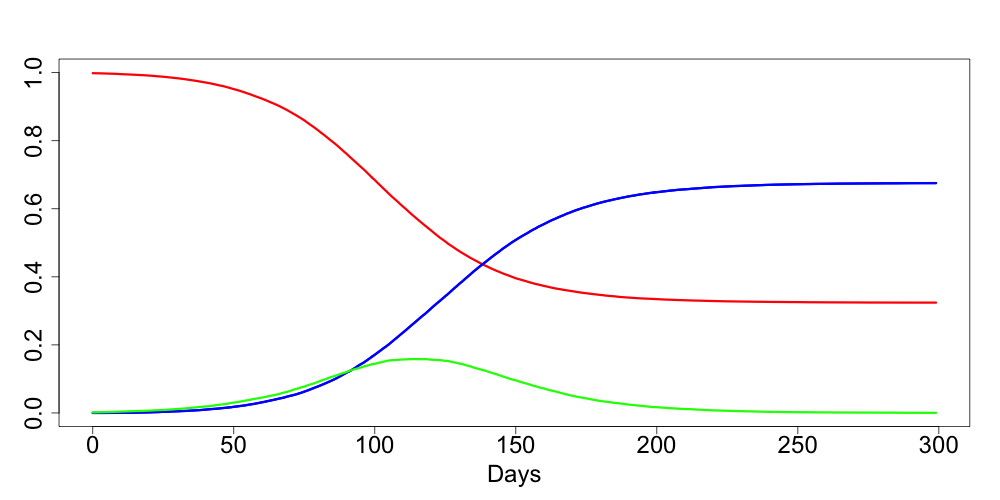}
  \caption{An uninterrupted run of a COVID-19 epidemic in homogeneous conditions}
\label{figure1}
\end{figure}

To start with, in Figure~\ref{figure1} we consider an uninterrupted run of an epidemic with $R_0=2.5$ in a homogeneous (one-group) population.
The starting time of an epidemic is  unknown and is even hard to define as the first transmissions of the virus take random and perhaps  long times. We start plotting  the curves after 0.5\% of the population are  infected.

In red colour, in Figure~\ref{figure1} and all plots below we plot values of $S(t)/N$, the proportion of people non-infected by time $t$. In blue, we plot $R(t)/N$, the proportion of people recovered from the disease (or dead) and in green we plot our main quantity of interest which is $I(t)/N$, the proportion of infected people at time $t$. The values of $R(t)$ do not play any part in  modelling and are plotted for information only.

The duration of time while a person with a virus is infected is modelled by  Erlang distribution with shape parameter $k=3$ and mean 21.
The simulations are  flexible and  we can easily change these values and plot similar curves with updated data.
Despite the full recovery taking slightly longer, $1/\sigma=21$ days is a good estimation for the period of time when people with severe infection may require intensive care, ventilator etc. Also, we believe that this is a suitable distribution for the period of time when such people may die.

From the value $I(t)$, we can  estimate the distribution of the number of deaths, at time $t$, as follows:
first, since on average an infected person is considered as infected for 21 days, we compute $I(t)/21$. At each particular day, any person can die with the probability which is typical for the  chosen population or sub-population. If we consider the whole population, then we apply the mortality rate for the population. However, if $I(t)$ refers to a particular group only, then we should apply the corresponding coefficient for the group. In any case, the distribution for the number of deaths at time $t$ can be roughly considered as binomial with parameters $I(t)$ and the probability of `success' $r \sigma$, where $r$ is the mortality rate. Different authors disagree on the values of  mortality $r$ for the COVID-19; see, for example,~\cite{npr}. UK's experts believe $r\simeq 0.009$~\cite{Ferguson},  WHO sets the world-wide mortality rate at 0.034, the authors of \cite{Sweden} believe $r$ is very small and could be close to $0.001$, an Israeli expert D. Yamin  sets $r=0.003$, see \cite{haaretz}.

{\bf Example 1.} {\it Assume that the epidemic in Birmingham (with population size $N \simeq 1,086$m) was running  uninterrupted according to the scenario depicted in Figure~\ref{figure1}. The maximum value of $I(t)/N$ is $\simeq 0.163$ giving the maximum expected daily death toll of $0.163 \cdot 1086000 \cdot 0.009/21 \simeq 75.8$ assuming we use UK experts value $r= 0.009$. }

Although simple to implement, the estimation in Example 1 is likely to be wrong, as it would be for any heterogeneous city (e.g. London). Critically, as will be discussed in the next section, the maximum death toll { is likely to be} significantly lower than the above calculations suggest.

Summarizing,  the expected daily mortality curve  is simply a suitably scaled version of  $I(t)/N$ of Figure~\ref{figure1}.
The same is applied to the curves representing expected number of hospital beds and ventilators. The variability for the number deaths is naturally considerably higher than the variability for the numbers of required hospital beds and ventilators as the latter ones are correlated in view of the fact that each person occupies a bed (requires a ventilator) for a few days in a row but dies only once.

Unlike $I(t)$, the number of deaths $D(t)$ at each particular day is (approximately) known and potentially can be used for estimating the first and second derivatives of $I(t)/N$ and hence  the stage of the epidemic. It is, however, difficult to do for the following reasons:  randomness in the values $D(t)$, see above, and, more importantly, heterogeneity of large sub-populations, see next section.

One of the main targets of decision-makers for dealing with epidemics like COVID-19 can be referred to as `flattening the curve', where `the curve' is  $I(t)/N$ or any of its equivalents and 'flattening' roughly means  `suppressing the maximum'. We shall consider this in the next sections.

\section{Spatial heterogeneity of the population and heterogeneity of epidemic development   in different areas}
\label{sec:het}

The purpose of this section is to demonstrate that heterogeneity of epidemic developments for different sub-populations has significant effect on
`flattening the curve'.

Already when this paper was completed, the authors learned about a preprint \cite{medrxiv}
 which uses SIR modelling to produce somewhat similar conclusions with more specifics for COVID-19 in the UK.
However, SIR modelling using  numerous age-classes and many equations for each age-class different across different
parts of the country  resulting in an astronomical number of parameters; this makes any sensitivity analysis
 to various parameters  uncheckable. We believe that a combination of  stochastic models like the present one with standard SIR-based  models  can help in significantly  reducing
the number of parameters while also adding more flexibility to a hybrid model.

Consider the following situation. Assume that we have a population consisting of $M$ sub-populations (groups) $G_m$ with  similar demographic and social characteristics and these  sub-populations   are subject to the same epidemic which has started at slightly different times. Let the sizes of sub-population $G_m$  be $N_m$  with $N_1+\ldots+N_m=N$. In all cases, the curves $I_m(t)/N_m$ are shifted in time versions of the curve
$I(t)/N$ of Figure~\ref{figure1}.

\begin{figure}[h]
\centering
\begin{minipage}{.45\textwidth}
  \centering
  \includegraphics[width=1\textwidth]{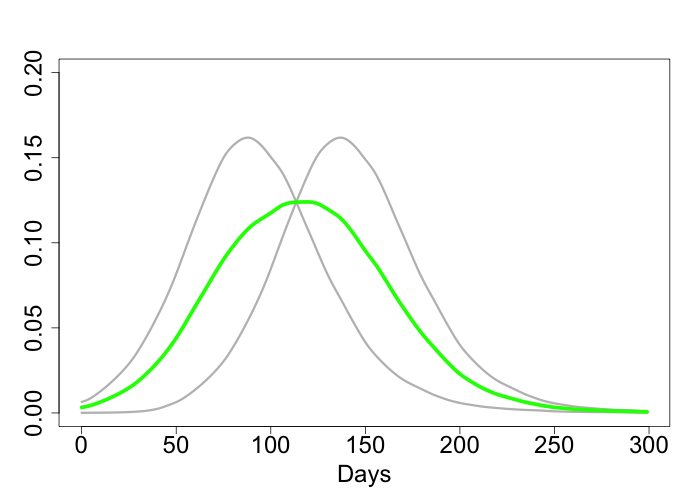}
\caption{$M=2$, $N_1=N_2$.}
\label{figure2}
\end{minipage}%
\begin{minipage}{.45\textwidth}
  \centering
\includegraphics[width=1\textwidth]{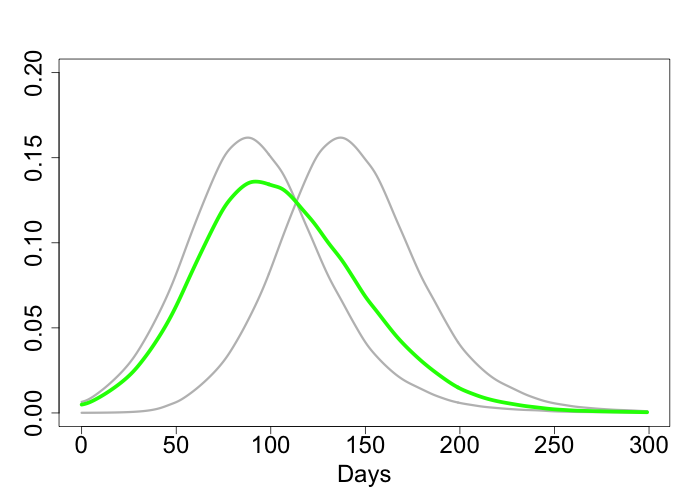}
\caption{$M=2$, $N_1=3N_2$}
\label{figure3}
\end{minipage}
\end{figure}

\begin{figure}[h]
\centering
\begin{minipage}{.45\textwidth}
  \centering
  \includegraphics[width=1\textwidth]{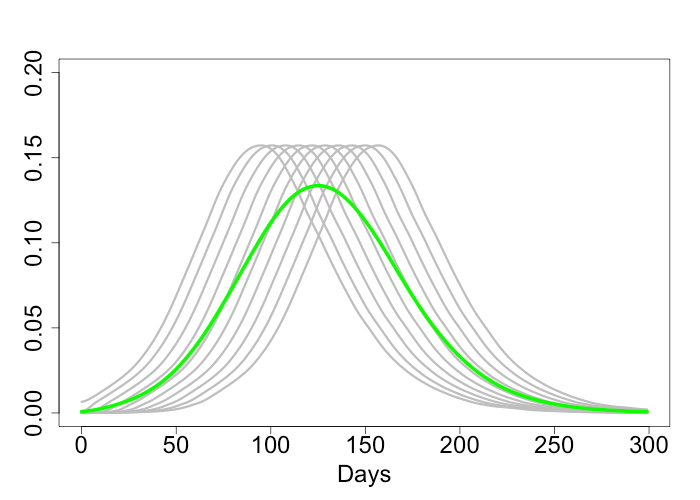}
\caption{$M=10$, $N_m=\frac{N}{10}$ ($m=1, \ldots,10$)}
\label{figure4}
\end{minipage}%
\begin{minipage}{.45\textwidth}
  \centering
\includegraphics[width=1\textwidth]{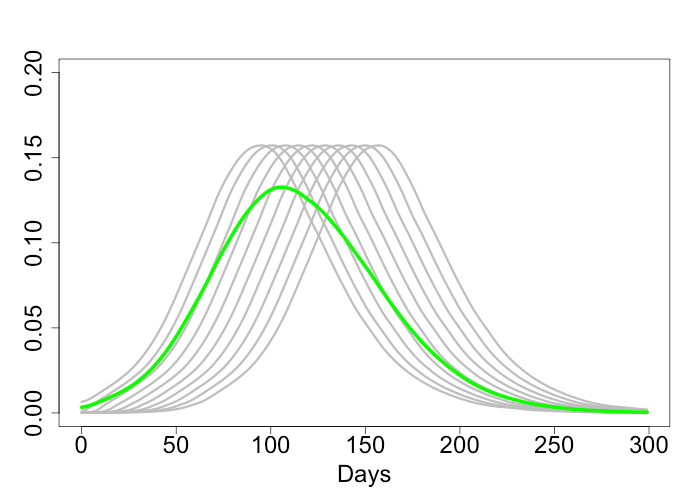}
\caption{$M=10$, $N_1=\frac{N}2 $, $N_m=\frac{N}{18}$ ($m>1$)}
\label{figure5}
\end{minipage}
\end{figure}

In Figures~\ref{figure2} and \ref{figure3}, $M=2$ and the second epidemic  started 50  days after the first one (incubation period is set to be 0).
In Figures~\ref{figure4} and \ref{figure5}, $M=10$ and each next epidemic cycle has started 7 days after the previous one. In all cases, we evidence the significant `flattening the curve' phenomenon. In Figures~\ref{figure2}--\ref{figure5}, the green line is used for the resulting curves $I(t)/N$.
The  maximal values of $I(t)/N$ in the examples depicted in Figures~\ref{figure2}--\ref{figure5} are 0.124, 0.136, 0.134 and 0.132, respectively.
This is significantly lower than the value 0.163 for the original curve. In the assumptions of Example 1, these would respectively lead to the maximum expected number of deaths   57.7, 63.3, 62.9 and 62.7.

Extra heterogeneity of different epidemics caused by the social and demographic heterogeneity of the sub-populations would further
`flatten the curve'. The following conclusions are  in line with what the Israeli expert D. Yamin states \cite{haaretz} and agree with the main conclusions of the paper \cite{Cirillo} which is rather critical towards standard epidemiological models.

{\bf Conclusions.} {\em (a) Isolation of sub-populations at initial stages of an epidemic is very important for preserving heterogeneity of the epidemic and  `flattenning the curve'; this flattening can be very significant.
(b) Epidemiological models based on the assumption of a homogeneous population but applied for populations consisting of  heterogeneous sub-populations may give completely misleading results. }

\section{An epidemic with intervention}
\label{sec:interv}


In this section, we assume that we make an intervention to the epidemic by introducing an isolation at a certain stage. Moreover, we assume
that isolation may be different  for 2 different groups. We define  a special group $G$ of more vulnerable people consisting, for example,
of all people  aged $70+$.
We define $\alpha=n/N$, where $N$ is the total population size and $n$ is the size of this special group~$G$.

Let $t_0$ be the moment of time when  the isolation occurs. It is natural to define $t_0$ from the condition $S(t_0)/N=x$,
where, for example, $x=0.9  $.
For $t< t_0$, the virus has been transmitted to people uniformly so that, conditionally a virus is transmitted, the probability that it reaches
 a person from group $G$ is $\alpha$.
For $t \geq t_0$, the virus has been transmitted to people in such a way that, conditionally a virus is transmitted, the probability that it reaches  a person from $G$ is $p=c \alpha$ with $0<c\leq 1$.
Moreover, at time $t_0$ the value of $R_0$ may change to $R_0^\prime$ in view of self-isolation.
The parameters in this model are:
\begin{itemize}
  \item $\alpha=n/N$: relative size of the group $G$;
  \item $x \in (0,1)$ defines the start of the isolation strategy;
  \item $c=p/\alpha$ defines the strength  of isolation of the group $G$;
  \item $R_0$ (initial);
  \item $R_{0}^\prime $:  the
   reproductive number  after intervention for $t\geq t_0$; it defines the  strength  of the overall isolation.
\end{itemize}

We have run a large number of scenarios {coded using a combination of R and Julia \cite{bezanson2017julia}; the code is provided in Appendix.} In  Figures~\ref{figure7}--\ref{figure11a},  we illustrate a few of these scenarios. In all these scenarios, we have chosen
$R_0=2.5$ and $\alpha=0.2$. The results are  robust towards values of these parameters (subject to faster or slower rate of the epidemic in dependence on $R_0$). In Section~\ref{sec:nhs} we use $\alpha=0.132$ to illustrate some specific results; $\alpha=0.2$ can be considered either as a generic value or, in the contents of Section~\ref{sec:nhs}, as the relative size of the group of vulnerable people which is larger
than the group of people aged $70+$.

We would like to emphasise that there is still a lot of uncertainty on who is vulnerable. It is very possible that the long term effect of the virus
might cause many extra morbidities in the coming years coming from severe cases who recover. It is also possible that the virus will mutate and
the new strain might affect younger populations more than the current strain. These, and many more possibilities, have a non-negligible probability
of occurring, and they have devastating effects. In future models we will try to address these possibilities.

{To estimate the number of COVID cases at a given time $t$ is  difficult  \cite{Germ2}. This implies that at the time of making an
intervention only very rough guesses about  the value of $x$, which is crucial for the future development of the epidemic, can be made.}

We distinguish different strengths of isolation  by values  of parameters $c$ and $R_{0}^\prime $:
\bea
c&=& \left\{
  \begin{array}{ll}
    1 & \mbox{no extra separation for people in $G$} \\
    0.5 & \mbox{mild  extra separation for people in $G$}  \\
    0.25 & \mbox{strong  extra separation for people in $G$}\\
  \end{array}
\right.
\\R_{0}^\prime& =& \left\{
  \begin{array}{ll}
    R_0 & \mbox{no  social distancing requirement  for general public} \\
    1.5 & \mbox{social distancing but no self-isolation requirement for general public}  \\
   1 & \mbox{self-isolation requirement for general public}\\
  \end{array}
\right.
\eea

The values $c=0.5$ and $c=0.25$ mean that under the condition that a virus is  infecting a new person,  the probabilities  that this new person belongs to $G$ are $1/3$ and $1/5$ respectively.

Measuring the level of compliance in the population and converting this to simple epidemiological measures $c$ and $ R_0'$ is hugely complex
problem which is beyond the scope of this paper.

The lines and  respective colours in Figures~\ref{figure7}--\ref{figure11a} are as follows.
\begin{itemize}
\item Solid blue line: $R(t)/N$,  the proportion of people recovered from the disease (or dead)  in case of no intervention,
as in Figure~\ref{figure1}.
\item Dashed blue line: $R(t)/N$ in case of  intervention.
\item Solid black line: $R_G(t)/n$ the proportion of people from group $G$ recovered from the disease (or dead)  in case of intervention.

\item Solid red line:  $S(t)/N$, the proportion of people non-infected by time $t$ in case of no intervention, as in Figure~\ref{figure1}.
\item Solid orange line: $S_G(t)/n$, the proportion of people susceptible to the virus from group $G$ non-infected by time $t$ in case of intervention.
\item Solid green line:  $I(t)/N$, the proportion of infected people at time $t$ with no intervention, as in Figure~\ref{figure1}.

\item Dashed green line:  $I(t)/N$, the proportion of infected people at time $t$ with  intervention.
\item Solid purple line: $I_G(t)/n$, the proportion of infected people from group $G$ at time $t$ with  intervention.
\end{itemize}

\begin{figure}[h]
\centering
  \includegraphics[width=0.85\linewidth]{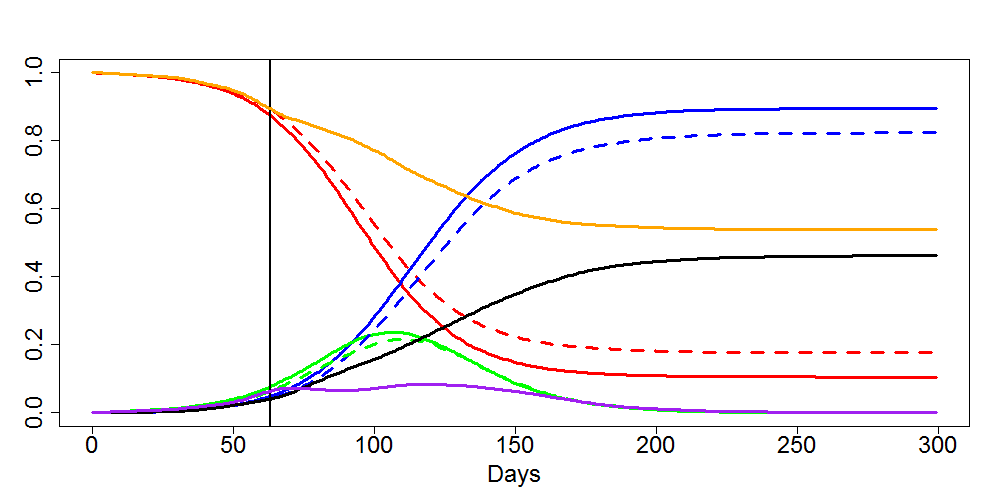}
  \caption{$x=0.9$, $c=0.25$, $R_0^\prime = 2.5$}
   \label{figure7}
\end{figure}

Considerably important, in view of the discussions of the next section, is  Figure~\ref{figure7}. In this figure,  for the initial data of Figure~\ref{figure1}, we strongly separate group $G$ at the time when 10\% of the population is infected. We make no call to the general public for social distancing. In this scenario, the intervention does not considerably change the proportion of  infected in the total population but it significantly `flattens the curve' for the group $G$: compare the purple and green lines. The maximum  of $I_G(t)/n$ is 0.086 which is 2.5 times lower  than 0.216, the maximum of $I(t)/N$. Note also the fact that the curve $I_G(t)/n$ is rather flat for  long time, during the main stage of development of the epidemic. Just after the peak of the  epidemic in the whole population, $I_G(t)/n$ peaks; it is caused by the presence of very large number of infected people from the general population.

\begin{figure}[h]
\centering
  \includegraphics[width=0.85\linewidth]{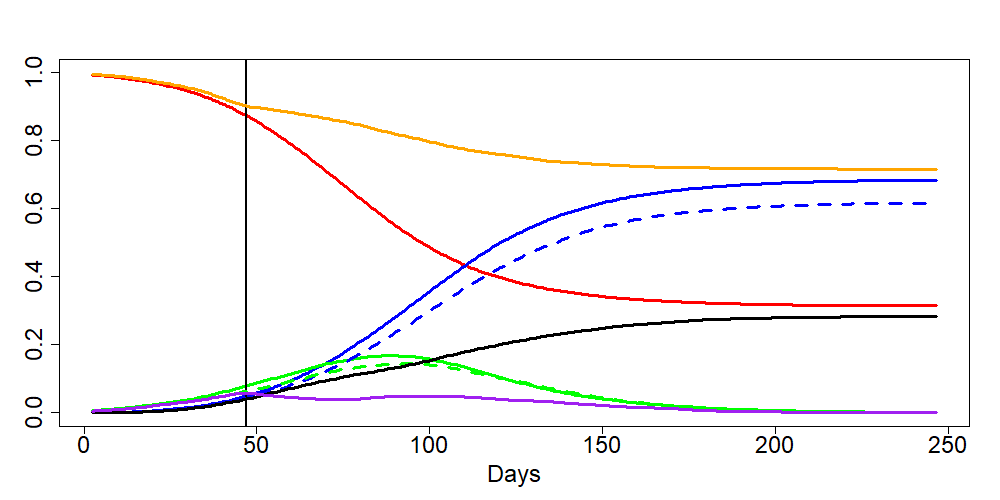}
  \caption{$x=0.9$, $c=0.25$, $R_0^\prime= 2.5$, no incubation period}
  \label{figure7a}
\end{figure}

The scenario { which} led to Figure~\ref{figure7} is important and hence it has been re-run for a few  variations of the model. Figure~\ref{figure7a} shows the results of the simulations where we have removed the incubation period. The maximum of $I_G(t)/n$ is now 0.075 which is more than twice lower than 1.63, the maximum of $I(t)/N$. Note that in the scenario with no incubation period the whole epidemic is milder as the virus lives longer.

In  Figure~\ref{figure6} we use similar scenario as for Figure~\ref{figure7} but
we  separate group $G$ only mildly at the time when 10\% of the population is infected. We make no call to general public for social distancing. The curve $I_G(t)/n$ is  `flattened' for the group $G$ but in a considerably smaller degree than in the case of strong separation of $G$.

\begin{figure}[h]
\centering
  \includegraphics[width=0.85\linewidth]{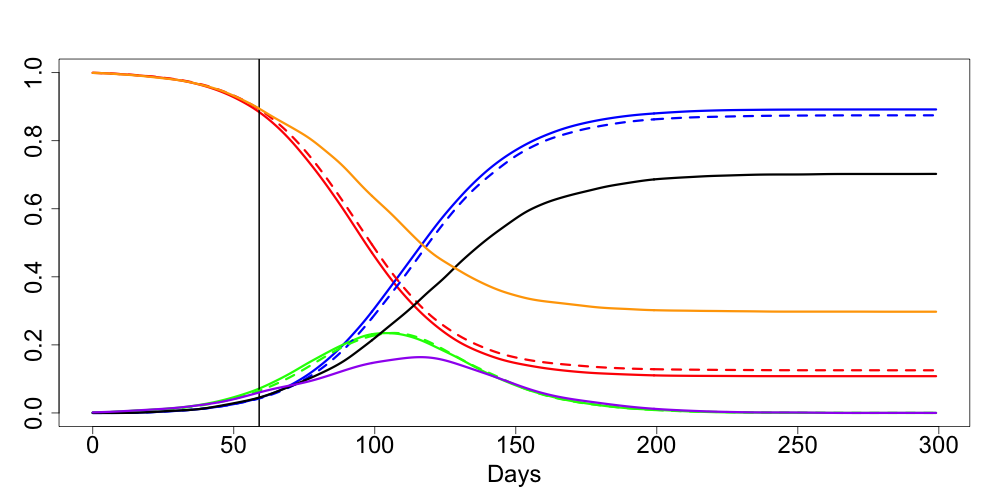}
  \caption{$x=0.9$, $c=0.5$, $R_0^\prime = 2.5$}
   \label{figure6}
\end{figure}

%

%

Figures~\ref{figure11c} and \ref{figure11b} illustrate the situation with mild and strong separation of people from $G$ complemented with introduction of the social distancing for general public. Interestingly enough, the effect of social distancing for general public gives less benefit than even mild isolation of people from group~$G$.

Figure~\ref{figure11a}
 illustrates the scenario  with no call to the general public for social distancing but with strong separation of the people from $G$ at
at a later stage of epidemic,  when 20\% of the population is infected. The effect is  similar to the one observed in Figures~\ref{figure7} and~\ref{figure7a} except for the fact that the call for isolating the group $G$ came slightly late.

\begin{figure}[h]
\centering
  \includegraphics[width=0.85\linewidth]{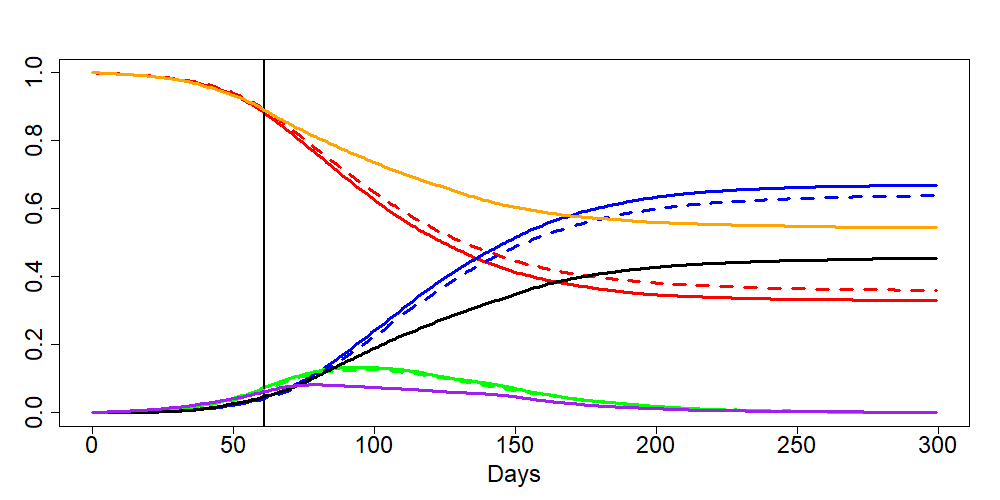}
  \caption{$x=0.9$, $c=0.5$, $R_0^\prime = 1.5$}
   \label{figure11c}
\end{figure}
\clearpage

\begin{figure}[h]
\centering
  \includegraphics[width=0.85\linewidth]{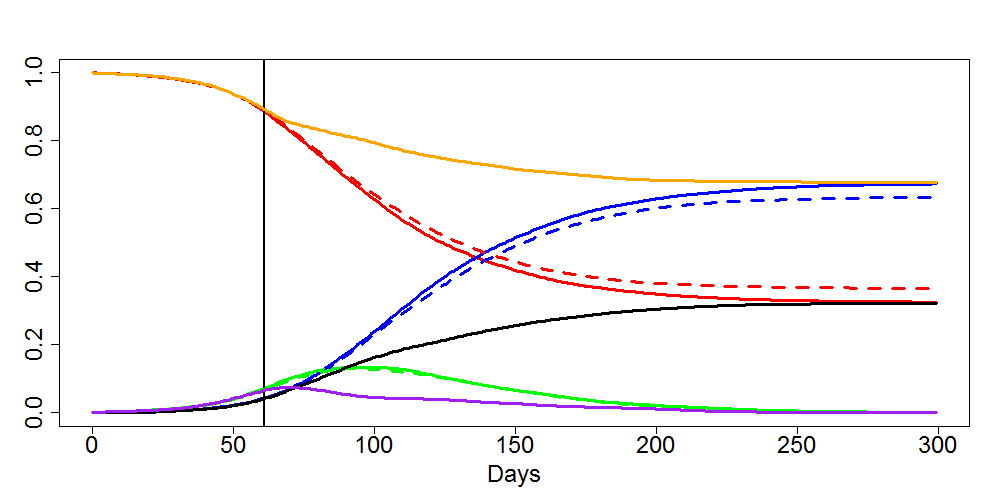}
  \caption{$x=0.9$, $c=0.25$, $R_0^\prime = 1.5$}
   \label{figure11b}
\end{figure}


\begin{figure}[h]
\centering
  \includegraphics[width=0.85\linewidth]{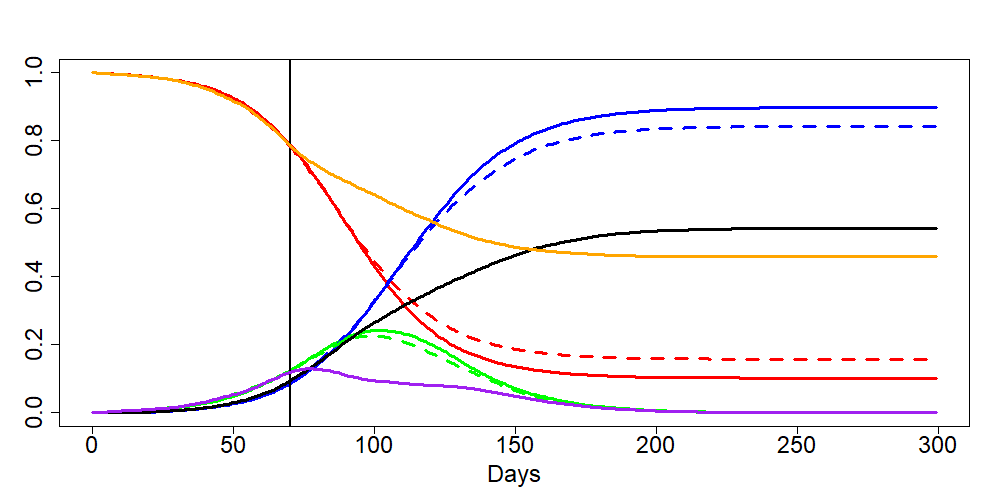}
  \caption{$x=0.8$, $c=0.25$, $R_0^\prime = 2.5$}
   \label{figure11a}
\end{figure}


{\bf Conclusion.}
{\em By considering a number of scenarios we have observed  an extreme sensitivity of the negative
 consequences of the epidemic to the degree of separation of vulnerable
people. Sensitivity to the parameter measuring the degree of self-isolation for the whole population is less apparent although it is much costlier.}

{This conclusion is very much in line with  recommendations of leading German epidemiologists
\cite{Germ1}.
}

%

%
%

%
%
%

%
%
%

%
%
%

%
%
%
%
%

\section{Consequences for the mortality and impact on NHS}
\label{sec:nhs}

Results of the type presented in Figures~\ref{figure7}--\ref{figure11a} can be translated into the language of the expected number of death and expected number of beds required. To do this we extend the observations we made at the end of Section~\ref{sec2} concerning translation of the curve $I(t)/N$ into the curves  for the expected number of death $ED(t)$ and expected number of hospital beds in the UK. We use the common split of the UK population into following age groups:
\bea
\;G_1=[0,19],\;\; G_2=[20,29],\; G_3=[30,39],\;
G_4=[40,49],\; \\G_5=[50,59], \;G_6=[60,69], \;G_7=[70,79], \;G_8=[80,\infty)
\eea
and corresponding  numbers $N_m$ ($m=1, \ldots, 8$; in millions) taken from \cite{statista}
$$[15.58, 8.71, 8.83, 8.50, 8.96, 7.07, 5.49, 3.27] \;\; {\rm with} \; N=66.41$$

The death probabilities $p_m$ are given from Table 1 in \cite{Ferguson} and replicated many times by the BBC and other news agencies are:
\bea
\label{eq:prob}
[ 0.00003, 0.0003, 0.0008, 0.0015, 0.006, 0.022, 0.051, 0.093]\, .
\eea
Unfortunately, these numbers do not match the other key number given in \cite{Ferguson}: the UK average mortality rate which is estimated to be  about $0.9\%.$
As we feel the value of the  UK average mortality rate is more important, we have multiplied all  probabilities above by
0.732 to get the average mortality rate to be  $0.9\%.$

Defining the group $G$ as a union of groups $G_7$ and $G_8$, we have $n=5.49+3.27= 8. 76$m and $\alpha= 8.76/66.41 \simeq 0.132$.
We then compute the mortality rate in the group $G$ and for the rest of population  by
$$
r_G= \frac{N_7 p_7+N_8 p_8}{n}\simeq 0.049; \;\; r_{other}= \frac{N_1 p_1+\ldots+ N_6 p_6}{N-n }\simeq  0.0030. \;\;
$$

For estimating the average number of death in the group $G$ and for the rest of population we can the use the formulas
\bea \label{eq:bb}
ED_G(t)=r_G \sigma I_G(t) \;\; {\rm and } \;\; ED_{other}(t)=r_{other} \sigma (I(t)-I_G(t))
\eea
respectively. The average numbers of hospital beds required for two different groups are proportional to these numbers.

We have run a series of  scenarios for  an UK epidemic without taking into account spatial and social heterogeneity of the society
assuming that we would have separated (mildly and strongly) the group $G$ of $70+$ people at the time $t_0$ when  10\% of the population were infected with no call  to general public for social distancing. We marked $t_0$ as March 23, 2020.

The two  scenarios (with $c=0.5$ and $c=0.25$) we have used for illustrating this technique are different from the scenarios used for plotting Figures~\ref{figure7} and
\ref{figure7a} only by the value of $\alpha$. For Figures~\ref{figure7} and
\ref{figure7a}, we have chosen $\alpha=0.2$ but for the group $G$ of $70+$ old people in the UK we have
$\alpha= 0.13$.

\begin{figure}[h]
\centering
\begin{minipage}{.45\textwidth}
  \centering
  \includegraphics[width=1\textwidth]{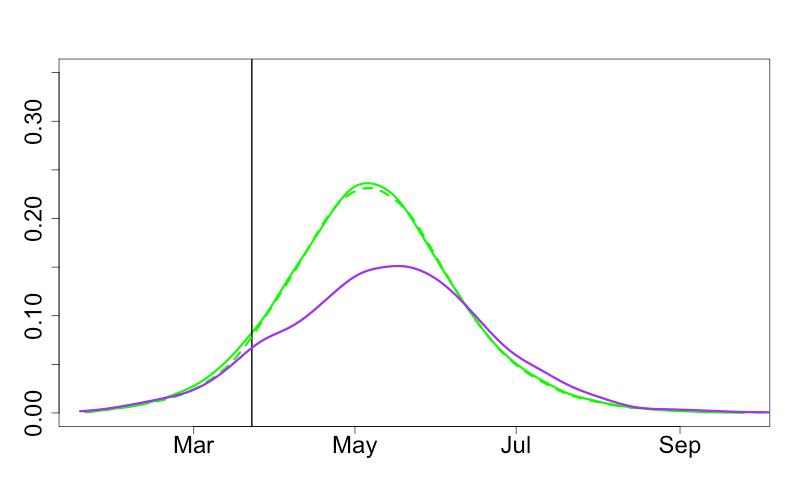}
\caption{$c=0.5$}
\label{figure1a}
\end{minipage}%
\begin{minipage}{.45\textwidth}
  \centering
\includegraphics[width=1\textwidth]{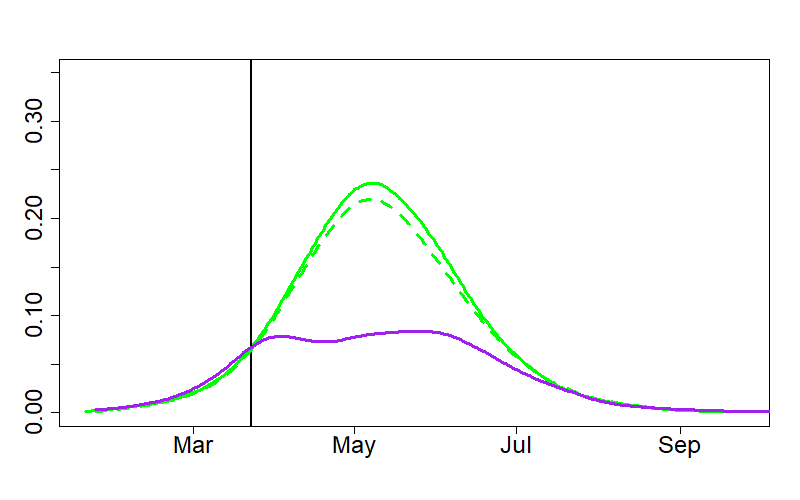}
\caption{$c=0.25$ }
\label{figure1b}
\end{minipage}
\end{figure}

In Figures~\ref{figure1a} and~\ref{figure1b} we plot (using the same colours as in  Figures~\ref{figure7} and \ref{figure7a}) the following curves:
$I(t)/N$ (solid green, no intervention on March 23), $I(t)/N$ (dashed green,  intervention on March 23) and $I_G(t)/n$ (purple,
 intervention on March 23).

\begin{figure}[h]
\centering
\begin{minipage}{.45\textwidth}
  \centering
  \includegraphics[width=1\textwidth]{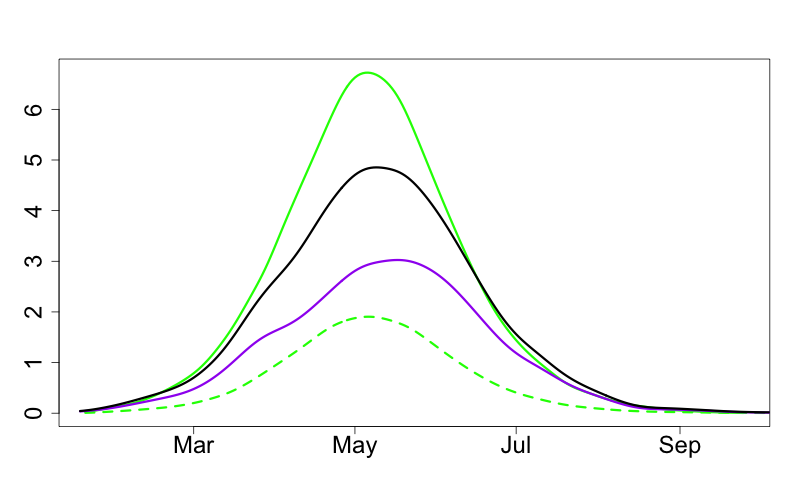}
\caption{$c=0.5$}
\label{figure1c}
\end{minipage}%
\begin{minipage}{.45\textwidth}
  \centering
\includegraphics[width=1\textwidth]{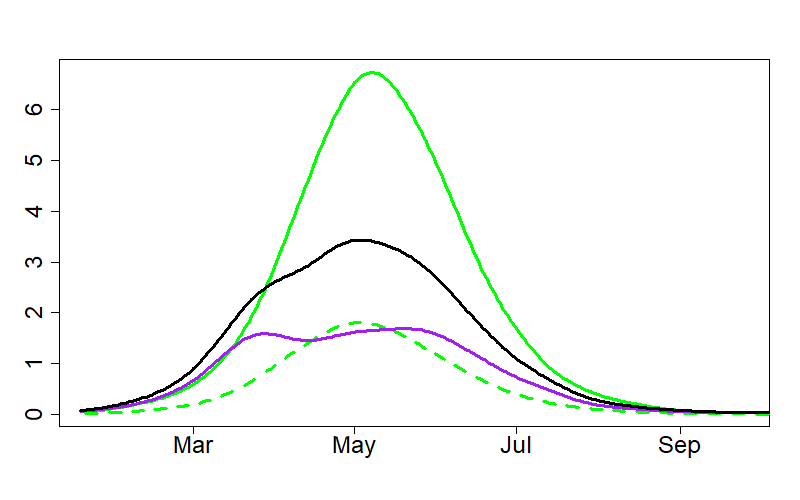}
\caption{$c=0.25$ }
\label{figure1d}
\end{minipage}
\end{figure}

 In Figures~\ref{figure1c} and~\ref{figure1d} we plot the curves for the estimated average number of deaths:
 $ED(t)$  (solid green, no intervention on March 23), $ED_G(t)$ (purple,  intervention on March 23), $ED_{other}(t)$ (dashed green,  intervention on March 23) and combined $E[D_G(t)+D_{other}(t)]$ (black,  intervention on March 23).

 We can deduce from  Figure
 \ref{figure1d} (taking into account the extra factor of hospital beds availability) that strong separation of $70+$ old people alone would have
 reduced the expected number of death by at least the factor of 2. Another  feature of the scenario with $c=0.25$ is a roughly 50/50 split between the number of deaths in the $70+$ and $70-$ groups.

The y-axis in Figures~\ref{figure1c} and~\ref{figure1d}, after multiplication by 140,  can be  roughly interpreted as hundreds in the London epidemic assuming $R_0=2.5$, $x=0.9$ on March 23  and homogeneity of the epidemic (as there is about 9m people in London out of 64.1m).
As mentioned in Section~\ref{sec:het}, deaths numbers in other regions of the UK should be expected to be lower.

\begin{figure}[h]
\centering
\begin{minipage}{.45\textwidth}
  \centering
  \includegraphics[width=1\textwidth]{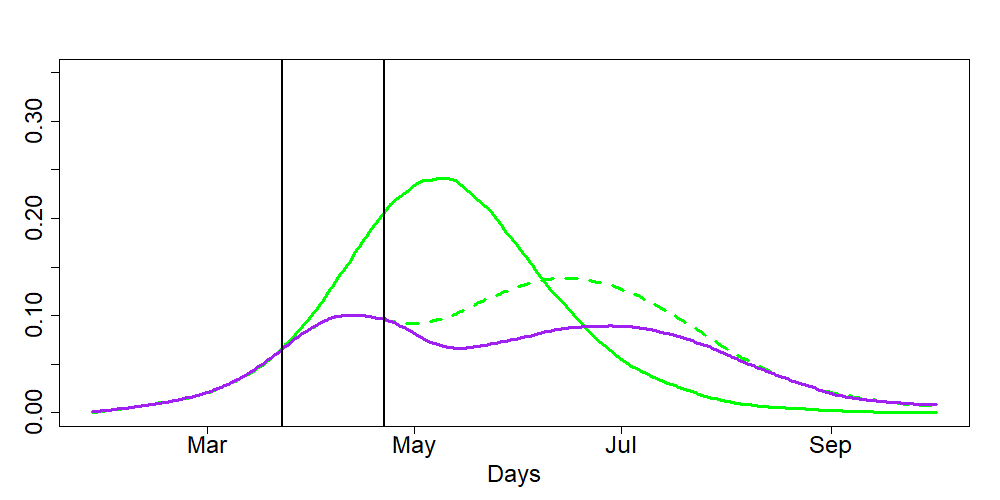}
\caption{$c=0.5$}
\label{figure_new1a}
\end{minipage}%
\begin{minipage}{.45\textwidth}
  \centering
\includegraphics[width=1\textwidth]{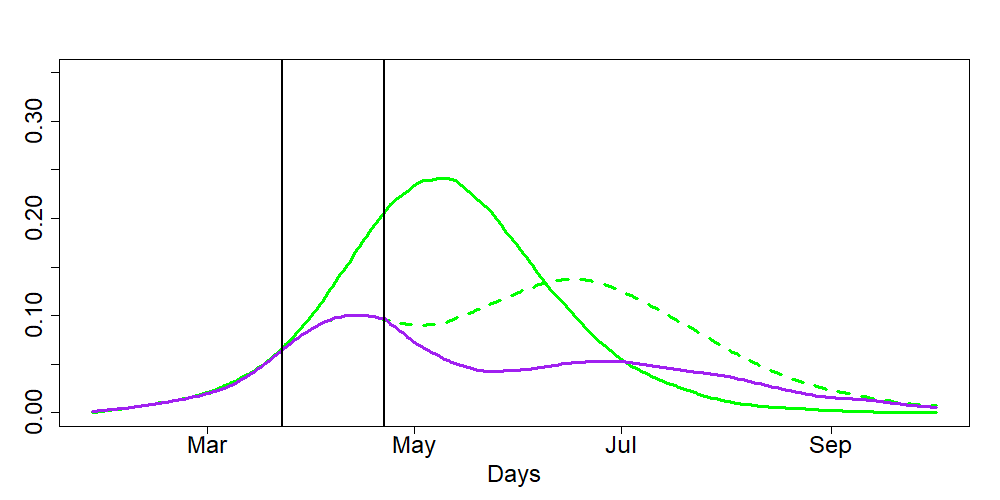}
\caption{$c=0.25$ }
\label{figure_new1b}
\end{minipage}
\end{figure}

\begin{figure}[h]
\centering
\begin{minipage}{.45\textwidth}
  \centering
  \includegraphics[width=1\textwidth]{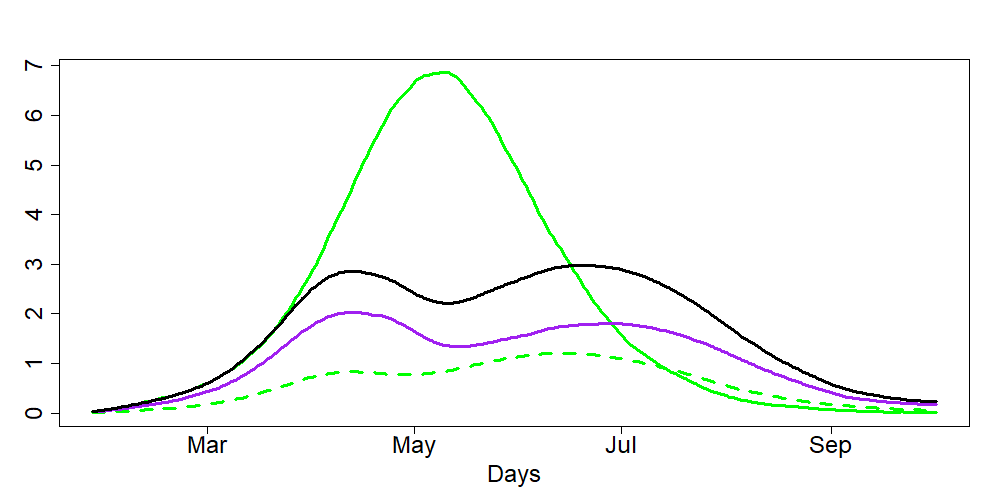}
\caption{$c=0.5$}
\label{figure_new1c}
\end{minipage}%
\begin{minipage}{.45\textwidth}
  \centering
\includegraphics[width=1\textwidth]{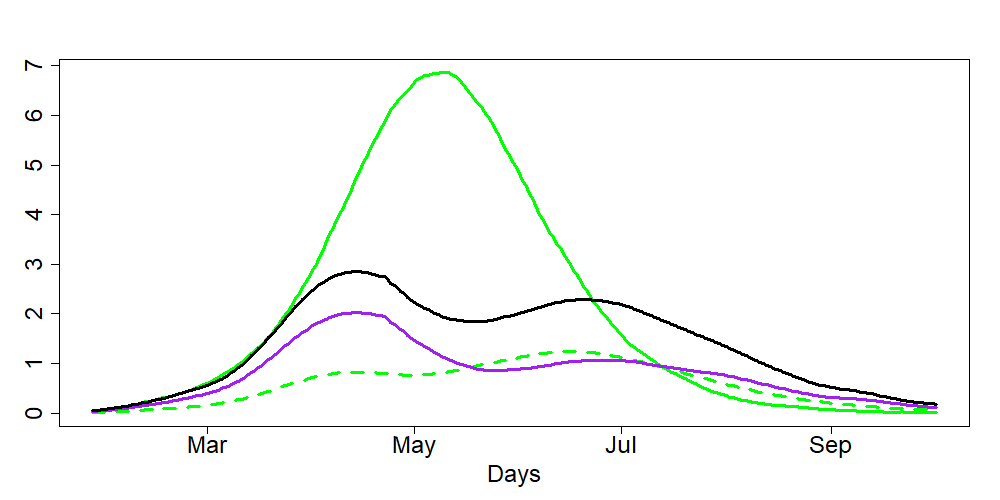}
\caption{$c=0.25$ }
\label{figure_new1d}
\end{minipage}
\end{figure}

We then have run through the  scenarios
with partial lockdown on March 23 reducing $R_0=2.5$ to  $R_0^\prime=1$ and with a return, 30 days later, to  $R_0=2.5$
with $c=0.5$ and $c=0.25$.
Results are plotted in Figures~\ref{figure_new1a}-\ref{figure_new1d}. The style of Figure is exactly the same as for Figures \ref{figure1a}-\ref{figure1d}.
If the value $x$ (proportion of non-infected people on March 23) happens to be larger than $0.9$ then the second wave of epidemic should be expected to  be (perhaps, considerably) larger. If $x<0.9$ then the second wave will be less pronounced.

\begin{figure}[h]
\centering
\begin{minipage}{.45\textwidth}
  \centering
  \includegraphics[width=1\textwidth]{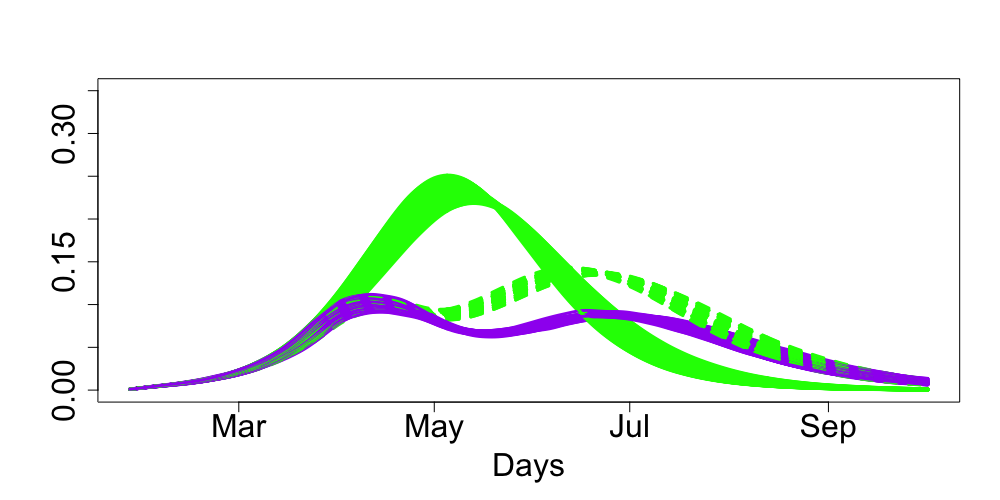}
\caption{$c=0.5$, $k \in [2,4]$}
\label{figure_new5a}
\end{minipage}%
\begin{minipage}{.45\textwidth}
  \centering
\includegraphics[width=1\textwidth]{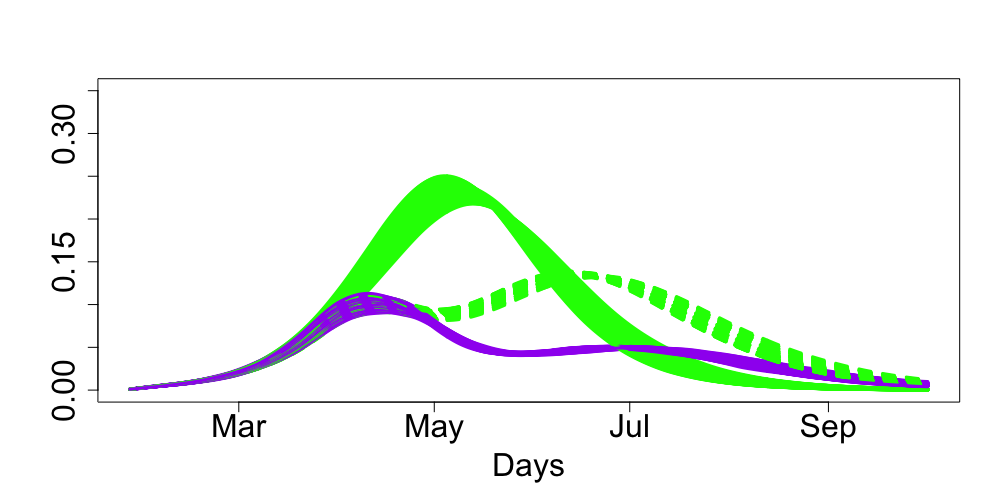}
\caption{$c=0.25$, $k \in [2,4]$ }
\label{figure_new5b}
\end{minipage}
\end{figure}

All these figures are given for illustration only without any claim on accurate predictions as  there is an uncertainty of the outputs towards the choice of several parameters describing the virus characteristics but the sensitivity of the model towards the choice of these parameters is not yet adequately assessed. {Figures~\ref{figure_new5a}-\ref{figure_new5b}
 illustrate sensitivity of the scenario of Figures~\ref{figure_new1a}-\ref{figure_new1b}    with respect to the choice of $k$, {the shape parameter of the Erlang distribution defined in Section~\ref{sec2}.} In these figures, we plot 100 trajectories similar to the trajectories of Figures~\ref{figure_new1a}-\ref{figure_new1b}   with the values of $k$ taken at random in the interval $[2,4]$.}

\smallskip

\pagebreak

{\bf Conclusion.} {\em The outputs of the model developed above can be translated into the language of the expected number of death due to the epidemic. This language amplifies the  findings of the previous section stating that isolation of a relatively small percentage of population
may hugely reduce the death toll of the epidemic. }

\section*{Some of the next steps and open problems}

\begin{itemize}

   \item Run many different scenarios to get better understanding of the current situation with the epidemic and what can be done to effectively control it.
  \item Continue calibrating the model against existing epidemiological models and new data as it  emerges.
  \item By using methods of stochastic global optimization \cite{Anatoly1,Anatoly2}, teach the model to  learn from the data emerging  daily, for adapting  values of parameters describing the virus and hence the course of the epidemic.
\item Incorporate into the model algorithms for quickest on-line  detection of changes \cite{Valia,Tartakovsky} to learn about temporal and spatial heterogeneity of the development of the epidemic.
    \item Understanding the reproductive number $R_0$ in dependence on the population density in local areas and hence the mixing distribution for $R_0$ needed to combine sub-populations.
     \item Understanding  $R_0$ and other   virus parameters as random variables due to mutations.
     \item   Understanding the risk as a function depending on different factors  such as age and social groups.
        \item 	Use of OR models for studying NHS capacity issues  under different scenarios and correlated to the percentage of health providers getting infected by the virus.
  \item Properly quantify uncertainty for model predictions.
   \item Full-scale  sensitivity analysis  to different parameters.
  \item With better understanding of the role of parameters, formulate inverse problems like finding the stage of an epidemic by its early development.
  \item Use game theory \cite{Myerson} to produce tools for optimal decision making  with respect to slowing down the process of epidemic by enforcing spatial isolation and isolation of different sub-populations.
   \item Develop a more sophisticated model combining  stochastic differential calculus \cite{Watanabe}, fractional diffusion \cite{leonenko2017heavy} and  elements of direct simulation (the models of fractional diffusion will be used to understand the effects of slowing down  of the epidemic).
 \item Understand and quantify the differences in epidemic developments across different countries such as UK, USA, Italy and Spain.
 \item Studying long-term effects of coronovirus and future mortality
from Covid-19.
 \item Studying mortality from
other diseases due to the capacity/fragility of the health system during the epidemic.
\item Modelling  epidemic control strategies based on  testing,  tracing and isolation.
 \end{itemize}

\pagebreak

\section*{Appendix: Julia/R code used for computing the scenarios  of Section~\ref{sec:interv}}
\begin{lstlisting}
#### Main script with R and Julia ####

## Running code will require installation of Julia
##install.packages("JuliaCall")
library('JuliaCall')
## In the below, specify where Julia is located
julia <- julia_setup(JULIA_HOME= "/Applications/Julia-1.4.app/Contents/Resources/julia/bin")
## Install required packages in Julia and import into R
julia_library("Random")
julia_library("Distributions")

## Main simulation

## Set total time to run simulation
Total_time=250

## Below is the main procedure. Refer to paper for description of what parameters are.

julia_command("function Main_procedure(Time,R,R_prime,N,c,quantile,alpha,sigma)
              Time=Int(Time)
              N=Int(N)
              
              ## Set parameters for study
              beta = R*sigma              ## Transmission coefficient
              k=3                         ## parameter for Erlang Distribution
              alpha_prime = c*alpha       ## New alpha
              beta1 =R_prime*sigma        ## New transmission coefficient if R0 changes
              Infectious_parameter=1/7    ## Average of incubation period
              
              ## indexes of columns in the matrix 'Population'
              person=1
              age=2
              Group=3
              Susceptible=4
              Infected=5
              Recovered=6
              Time_til_infected=7
              time_of_next_inf=8
              time_of_recovery=9
              previous_infection=10
              
              ## Create initial population
              
              G1_split = Int(N*(1-alpha))   ## Number of people in first group
              G2_split = N-G1_split         ## Number of people in second group
              
              Population =zeros(Int32,N,10)
              Population[:,person]=[1:N;]
              Population[:,Susceptible].=1
              Population[1:G1_split,Group].=1
              Population[G1_split+1:N,Group].=2
              
              Num_of_ran = Int(round(N*0.005))
              ran=randperm(N)[1:Num_of_ran]
              qq=(1:Num_of_ran)/(Num_of_ran+1)
              temp1=floor.(cquantile.(Exponential(1/Infectious_parameter), qq))[randperm(Num_of_ran)]
              temp1[temp1.==0].=1
              Population[ran,Time_til_infected]=temp1  ## Time they will become infected and can pass on virus
              #Population[ran,]$Infected=0
              Population[ran,Susceptible].=1
              temp = floor.(cquantile.(Exponential(1/beta), qq))[randperm(Num_of_ran)] 
              temp[temp.==0].=1
              Population[ran,time_of_next_inf] = Population[ran,Time_til_infected] + temp  ## Time they pass on virus to someone
              temp2 = floor.(cquantile.(Gamma(k,1/(k*sigma)), qq))[randperm(Num_of_ran)]
              Population[ran,time_of_recovery] = Population[ran,Time_til_infected] + temp2 # Time this person recovers when infected
            


              ## Create variables for tracking
              
              Susceptible_track       = zeros(Time)     ## vector created to track those susceptible in whole population
              Infected_track          = zeros(Time)     ## vector created to track those infected in whole population
              Recovered_track         = zeros(Time)     ## vector created to track those recovered in whole population
              Second_group_track      = zeros(Time)     ## vector created to track those recovered in vulnerable population
              Second_group_track_susc = zeros(Time)     ## vector created to track those susceptible in vulnerable population
              Second_group_track_Inf  = zeros(Time)     ## vector created to track those infected in vulnerable population
              
              ## Begin simulation
              
              time_to_start_isolation=Time
              for t in 1:Time
              ##Isolation code. Only initialise isolation when proportion of suceptible people = quantile
              
              if (sum(Population[:,Susceptible])/N < quantile) && (t< time_to_start_isolation)
              time_to_start_isolation=t
              end
              
              ## Index the current people who have the chance to infect others.
              
              Current = Population[(Population[:,Time_til_infected].<t) .& (Population[:,Time_til_infected].>Int(0)) .& (Population[:,Recovered].==Int(0)),1]
              
              for i in Current
              Population[i,Infected] = 1               ## Person now becomes infected
              Population[i,Susceptible] = 0            ## Remove susceptible flag
              Population[i,previous_infection] = 1     ## Initialise previous infection flag
              
              ## If it is time to infect and not time to recover proceed
              
              if (Population[i,time_of_next_inf]==t & t<Population[i,time_of_recovery] )
              
              if ( Population[i,Group]==1  )
              
              ## Before isolation. We take alpha and not alpha prime
              if (t<time_to_start_isolation)
              X = ifelse(rand()<alpha,2,1)   ## Sample from which group to infect
              ## Randomly select one person from that group
              Y = Int(ifelse(X==1,ceil(G1_split*rand()),G1_split+ceil(G2_split*rand())))
              ## Check if person has virus previously. Otherwise give virus
              if ( Population[Y,previous_infection]==0 & Population[Y,Time_til_infected]==0 )
              temp = floor(rand(Exponential(1/beta)))
              Population[Y,Time_til_infected]= t+ floor(rand(Exponential(1/Infectious_parameter)))
              Population[Y,time_of_next_inf] = Population[Y,Time_til_infected]+ ifelse(temp==0,1,temp)
              Population[Y,time_of_recovery] = Population[Y,Time_til_infected]+ floor(rand(Gamma(k,1/(k*sigma))))
              end
              ## isolation initialised. We take alpha prime here
              else
              X = ifelse(rand()<alpha_prime,2,1)   ## Sample from which group to infect
              ## Randomly select one person from that group
              Y = Int(ifelse(X==1,ceil(G1_split*rand()),G1_split+ceil(G2_split*rand())))
              ## Check if person has virus previously. Otherwise give virus
              if ( Population[Y,previous_infection]==0& Population[Y,Time_til_infected]==0 )
              temp = floor(rand(Exponential(1/beta1)))
              Population[Y,Time_til_infected]= t+ floor(rand(Exponential(1/Infectious_parameter)))
              Population[Y,time_of_next_inf] = Population[Y,Time_til_infected]+ ifelse(temp==0,1,temp)
              Population[Y,time_of_recovery] = Population[Y,Time_til_infected]+ floor(rand(Gamma(k,1/(k*sigma))))
              end
              end
              
              
              ## Repeat the same above for Group 1 but for Group 2 (Vulnerable group)
              
              elseif ( Population[i,Group]==2  )
              
              if (t<time_to_start_isolation)
              X = ifelse(rand()<alpha,2,1)
              Y = Int(ifelse(X==1,ceil(G1_split*rand()),G1_split+ceil(G2_split*rand())))
              if ( Population[Y,previous_infection]==0& Population[Y,Time_til_infected]==0  )
              temp = floor(rand(Exponential(1/beta)))
              Population[Y,Time_til_infected]= t+ floor(rand(Exponential(1/Infectious_parameter)))
              Population[Y,time_of_next_inf] = Population[Y,Time_til_infected]+ ifelse(temp==0,1,temp)
              Population[Y,time_of_recovery] = Population[Y,Time_til_infected]+ floor(rand(Gamma(k,1/(k*sigma))))
              end
              else
              X = ifelse(rand()<alpha_prime,2,1)
              Y = Int(ifelse(X==1,ceil(G1_split*rand()),G1_split+ceil(G2_split*rand())))
              if ( Population[Y,previous_infection]==0 & Population[Y,Time_til_infected]==0  )
              temp = floor(rand(Exponential(1/beta1)))
              Population[Y,Time_til_infected]= t+ floor(rand(Exponential(1/Infectious_parameter)))
              Population[Y,time_of_next_inf] = Population[Y,Time_til_infected]+ ifelse(temp==0,1,temp)
              Population[Y,time_of_recovery] = Population[Y,Time_til_infected]+ floor(rand(Gamma(k,1/(k*sigma))))
              end
              end
              end
              

              ## If it is not time for the person infected to recover, find next time the person will transmit
              if (t<time_to_start_isolation)
              temp = floor(rand(Exponential(1/beta)))
              Population[i,time_of_next_inf]=Population[i,time_of_next_inf]+ ifelse(temp==0,1,temp)
              else
              temp = floor(rand(Exponential(1/beta1)))
              Population[i,time_of_next_inf]=Population[i,time_of_next_inf]+ ifelse(temp==0,1,temp)
              
              end
              
              ## If it is time to recover, set recovery flag =1 and remove infected flag.
              
              elseif ( Population[i,time_of_recovery]<t )
              Population[i,Recovered]=1
              Population[i,Infected]=0
              end
              end # for i in Current
              
              ## Update tracking variables
              
              index = t
              
              Susceptible_track[index]   = sum(Population[:,Susceptible])/N       ## Keep track of susceptible rate
              Recovered_track[index]     = sum(Population[:,Recovered])/N         ## Keep track of recovered rate
              Infected_track[index]      = sum(Population[:,Infected])/N          ## Keep track of infected rate
              Second_group_track[index]  = sum(Population[G1_split+1:N,Recovered])/G2_split   ## Keep track of recovered rate for vulnerable
              Second_group_track_susc[index] = sum(Population[G1_split+1:N,Susceptible])/G2_split ## Keep track of susceptible rate for vulnerable
              Second_group_track_Inf[index]  = sum(Population[G1_split+1:N,Infected])/G2_split  ## Keep track of Infected rate for vulnerable
              
              #global t=t+1
              end # for t
              
              Output     =  zeros(Time,8)
              Output[:,1] = Susceptible_track
              Output[:,2] = Recovered_track
              Output[:,3] = Infected_track
              Output[:,4] = Second_group_track
              Output[:,5] = Second_group_track_susc
              Output[:,6] = Second_group_track_Inf
              Output[:,7] .= time_to_start_isolation
              Output
              end")

# Set the parameters for the study
R=2.5
R_prime=2.5
N=1000000
c=1
quantile=0.8
alpha=0.2
sigma=1/21

########################
Scenario_1  =  julia_call("Main_procedure",Total_time,R,R_prime,N,c,quantile,alpha,sigma)
## Change c for different scenarios
c=0.25
Scenario_2  =  julia_call("Main_procedure",Total_time,R,R_prime,N,c,quantile,alpha,sigma)
########################
Scenario_1=t(Scenario_1)
Scenario_2=t(Scenario_2)

##  Plots for scenario 1
plot(seq(0,Total_time-1,1),Scenario_1[1,],type='l',col='red',ylim=c(0,1),lwd=3,lty=1,main='',xlab='Days',ylab='',cex.axis=2,cex.lab=2,cex.main=1.5)
lines(seq(0,Total_time-1,1),Scenario_1[2,],type='l',col='blue',lwd=3 ,lty=1  )
lines(seq(0,Total_time-1,1),Scenario_1[3,],type='l',col='green',lwd=3 ,lty=1  )
abline(v= Scenario_1[7,],lwd=2)

##  Lines for scenario 2
lines(seq(0,Total_time-1,1),Scenario_2[1,],type='l',col='red',ylim=c(0,1),lwd=3,lty=2,main='',xlab='Days',ylab='',cex.axis=2,cex.lab=2,cex.main=1.5)
lines(seq(0,Total_time-1,1),Scenario_2[2,],type='l',col='blue',lwd=3,lty=2)
lines(seq(0,Total_time-1,1),Scenario_2[3,],type='l',col='green',lwd=3,lty=2)
lines(seq(0,Total_time-1,1),Scenario_2[4,],type='l',col='black',lwd=3,lty=1  )
lines(seq(0,Total_time-1,1),Scenario_2[5,],type='l',col='orange',lwd=3,lty=1  )
lines(seq(0,Total_time-1,1),Scenario_2[6,],type='l',col='purple',lwd=3,lty=1  )





\end{lstlisting}

\bibliographystyle{unsrt}
\bibliography{Coronavirus}

\end{document}